\documentclass[%
 reprint,
 superscriptaddress,
 amsmath,amssymb,
 aps,
 prl
]{revtex4-2}

\usepackage{graphicx} % Required for inserting images
\usepackage{subcaption}
\usepackage{siunitx}
\usepackage{graphicx}% Include figure files
\usepackage{dcolumn}% Align table columns on decimal point
\usepackage{bm}% bold math
\usepackage{booktabs}       % professional-quality tables
\usepackage{cancel}
\usepackage{ulem}
\expandafter\let\csname equation*\endcsname\relax
\expandafter\let\csname endequation*\endcsname\relax
\usepackage{amsmath}
\usepackage{ragged2e}
\usepackage{cleveref}

\begin{document}

\preprint{APS}

%\title{The "Core" impact on electron-induced Shifts in $^{229}$Th nuclear transition frequencies
%\textcolor{red}{Accurate Prediction of $^{229}$Th Nuclear Transitions Difference through Energy Variation in Atomic Level.}}
\title{Prediction of Nuclear Clock Transitions Frequency Difference between $^{229}$Th$^{3+}$ and $^{229}$Th$^{4+}$ via \textit{ab-initio} Self-Consistent Field Theory}

\author{Ran Si}
\email{rsi@fudan.edu.cn}
\affiliation{Key Laboratory of Nuclear Physics and Ion-Beam Application (MOE), Institute of Modern Physics, Fudan University, 200433 Shanghai, China}

\author{Chaofan Shi}
\affiliation{Key Laboratory of Nuclear Physics and Ion-Beam Application (MOE), Institute of Modern Physics, Fudan University, 200433 Shanghai, China}

\author{Nan Xue}
\affiliation{Key Laboratory of Nuclear Physics and Ion-Beam Application (MOE), Institute of Modern Physics, Fudan University, 200433 Shanghai, China}
\affiliation{Research Center for Theoretical Nuclear Physics, NSFC and Fudan University, 200438 Shanghai, China}
%\affiliation{School of Nuclear Science and Technology,Lanzhou University, 730000 Lanzhou, China}

\author{Xiangjin Kong} 
\email{kongxiangjin@fudan.edu.cn}
\affiliation{Key Laboratory of Nuclear Physics and Ion-Beam Application (MOE), Institute of Modern Physics, Fudan University, 200433 Shanghai, China}
\affiliation{Research Center for Theoretical Nuclear Physics, NSFC and Fudan University, 200438 Shanghai, China}

\author{Chongyang Chen}
\affiliation{Key Laboratory of Nuclear Physics and Ion-Beam Application (MOE), Institute of Modern Physics, Fudan University, 200433 Shanghai, China}

\author{Bingsheng Tu}
\email{bingshengtu@fudan.edu.cn}
\affiliation{Key Laboratory of Nuclear Physics and Ion-Beam Application (MOE), Institute of Modern Physics, Fudan University, 200433 Shanghai, China}

\author{Yu-Gang Ma}
\email{mayugang@fudan.edu.cn}
\affiliation{Key Laboratory of Nuclear Physics and Ion-Beam Application (MOE), Institute of Modern Physics, Fudan University, 200433 Shanghai, China}
\affiliation{Research Center for Theoretical Nuclear Physics, NSFC and Fudan University, 200438 Shanghai, China}

\date{\today}% It is always \today, today,

\begin{abstract}

The $^{229}\text{Th}$ isotope is a promising candidate for nuclear clocks, with its transition frequency influenced by electron-induced nuclear frequency shifts. This effect is comparatively small and requires high-precision theoretical calculations. In this work, we employed a non-perturbative multi-configuration Dirac-Hartree-Fock (MCDHF) method,  in contrast to the perturbation theory used previously, to resolve the field shift effect. This method accounts for subtle differences in the nuclear potential while considering the $^{229}\text{Th}$ isotope in both its ground and isomeric states. Consequently, the nuclear transition frequency difference of between $^{229}\text{Th}^{3+}$ and $^{229}\text{Th}^{4+}$ was determined to be $-639$~MHz with computational convergency down to 1~MHz. 
%\textcolor{blue}{This effect is relatively small and is therefore typically estimated using perturbation theory. 
%To address the uncertainties associated with perturbative approaches, we employ a non-perturbative method in this study. 
%By performing four independent MCDHF calculations with our newly developed GRASPG code, we precisely determine the transition frequency difference between $^{229}\text{Th}^{3+}$ and $^{229}\text{Th}^{4+}$ to be $-639$~MHz. }
Given recent precision measured transition frequency of $^{229}\text{Th}^{4+}$in $^{229}\text{Th}$-doped CaF$_2$ [Nature 633, 63 (2024)], the transition frequency of isolated $^{229}\text{Th}^{3+}$ is predicted to be $2,020,406,745 (1)_\text{comp.}(77)_{\delta \langle r^2 \rangle} (100)_\text{ext.}$~MHz, with brackets indicating uncertainties stemming from our atomic structure computations, the input nuclear charge radii from nuclear data tables, and the influence of the crystal environment as reported in the literature. This provides valuable guidance for direct laser excitation of isolated $^{229}\text{Th}^{3+}$ based on ion traps experiments.
\end{abstract}

\maketitle

%\tableofcontents

%\section{Introduction}

The $^{229}\text{Th}$ isotope is currently considered the most promising candidate for the first nuclear clock, due to its low-lying isomer at about 8.4 eV above the ground state, which is accessible for precision laser spectroscopy~\cite{Beeks2021}. Beyond its high accuracy, nuclear clocks offer exceptional stability, as nuclear transitions are largely unaffected by external environments~\cite{EPeik_2003}. Furthermore, these clocks have the potential to advance fundamental physics~\cite{Peik_2021,RevModPhys.90.025008,PhysRevLett.97.092502,PhysRevA.102.052833,PhysRevLett.102.210801,PhysRevD.106.095031,PhysRevLett.115.201301,PhysRevD.91.015015}. There are two main approaches for constructing a $^{229}\text{Th}$ clock: solid-state systems, such as $^{229}\text{Th}$-doped crystals~\cite{Jackson_2009,Kazakov_2012,Dessovic_2014,Pimon_2020,Kraemer2023ObservationOT,Nature2024} and thin films~\cite{Zhang2024229ThF4TF}, and ion traps~\cite{PhysRevLett.106.223001,PhysRevLett.108.120802,PhysRevA.99.052514,Yamaguchi2024LaserSO}. Recently, the nuclear transition of $^{229}\text{Th}^{4+}$ in solid-state samples has been successfully excited using lasers~\cite{PhysRevLett.132.182501,PhysRevLett.133.013201,Nature2024}, while direct resonant excitation of $^{229}\text{Th}^{3+}$ in an ion trap has yet to be achieved.

For the photoexcitation of narrow nuclear transitions in experiments, a critical requirement is the precise tuning of the laser frequency to match the transition resonance energy, as scanning a broad range of possible resonant energies is impractical. Therefore, it is essential to predict the frequency of unobserved transitions as accurately as possible before conducting the experiment. In this context, the frequency difference of the nuclear transition between $^{229}\text{Th}^{3+}$ and $^{229}\text{Th}^{4+}$ is urgently needed to enable direct laser excitation of $^{229}\text{Th}^{3+}$ in an ion-trap system, particularly since the transition frequency in $^{229}\text{Th}^{4+}$ has already been measured with kHz precision~\cite{Nature2024}. Therefore, the impact on the nuclear transition frequency shift induced by the adding electron must be solved. 
A method for calculating the frequency difference in nuclear transitions between different charge states of $^{229}\text{Th}$ was presented in~\cite{PhysRevLett.131.263002}, which is analogous to the approach used for determining the isotope shift in atomic transition frequencies via the first-order perturbation theory. 
%In addition, assuming that the [Rn] core remains unchanged across all charge states of Th II, Th III and Th IV neglects the impact on the electronic core wave function induced by the valence electron. 

In this Letter, \textit{ab initio} calculations based on a non-perturbative method were conducted to determine the nuclear frequency difference between $^{229}\text{Th}^{3+}$ and $^{229}\text{Th}^{4+}$. By analyzing the energy variance of the atomic ground state in $^{229}\text{Th}^{3+}$, $^{229m}\text{Th}^{3+}$, $^{229}\text{Th}^{4+}$, and $^{229m}\text{Th}^{4+}$, the frequency shift of the nuclear clock transition in $^{229}\text{Th}^{3+}$ is determined to be 639 MHz lower than that in $^{229}\text{Th}^{4+}$. Given the recently measured high-precision transition frequency of $^{229}\text{Th}^{4+}$ in $^{229}\text{Th}$-doped CaF$_2$, our results represent a significant step toward accurately determining the yet-undetected transition frequency of isolated $^{229}\text{Th}^{3+}$, and provides valuable guidance for future laser excitation experiments in ion-trap systems.

Since all nuclei are not point-like charges, the Coulomb potential experienced by an electron—particularly in s-orbitals, which have a nonzero probability density at the nucleus—is slightly modified due to variations in the nuclear charge distribution. This effect, known as the field shift, induces a frequency shift in both atomic and nuclear transition and can be solved by the first-order perturbation theory~\cite{PhysRevA.31.2038},
\begin{equation}\label{Fsn}
\Delta E^{(1)}_{\text{FS}} = \sum_{n\geq 0, \text{even}} F_{n} \delta \langle r^{n+2} \rangle ^{A,A'},
\end{equation}
\noindent where $F_{n}$ is the field shift electronic factor~\cite{PhysRevA.31.2038} and $\delta \langle r^{n+2} \rangle ^{A,A'}$ is the nuclear parameter reflecting the isotopic variation of the nuclear charge distribution~\cite{PhysRev.188.1916}.
Typically, neglecting the nucleus deformation, only the first term, i.e. $F_{0}$ in Eq.~[\ref{Fsn}] is accounted for the calculation on field shift, and consequently, the nuclear transition energy difference between $^{229}\text{Th}^{3+}$ and $^{229}\text{Th}^{4+}$ can be determined as given in~\cite{PhysRevLett.131.263002},
\begin{equation}\label{FS}
\Delta E_{N}^{(1)} \approx ( F_{0,\rm ^{229}Th^{3+}} - F_{0,\rm ^{229}Th^{4+}} ) \delta \langle r^2 \rangle ^{229\text{m},229},
\end{equation}
\noindent where $ \delta \langle r^2 \rangle ^{229\text{m}, 229}$  is the difference in the rms nuclear charge radius of the isomeric and ground nuclear states,  $F_{0,\rm ^{229}Th^{3+}}$ and $F_{0,\rm ^{229}Th^{4+}}$ are the leading field shift electronic factors of the ground electronic state of $\rm ^{229}Th^{3+}$ and $\rm ^{229}Th^{4+}$, respectively.
The leading term of field shift electronic factor can be estimated by
\begin{equation}
F_{0}=\frac{2\pi}{3}Z | \Psi (\bm 0) |^2,
\end{equation}
\noindent where $ | \Psi (\bm 0)|^2$ is the total probability density of the electronic wave function at the origin, which can be estimated by taking the $r \rightarrow 0$ limit of the electron density $| \Psi (\bm {0})|^2 = \rm lim_{r \rightarrow 0} \rho^e ( \mathbf{r})$~\cite{NAZE20132187}.
%Additionally, the electronic wave function can vary depending on whether the nucleus is in its ground or isomeric state, as the nuclear charge distribution differs between these states, which causes a deviation in the calculation of $F_{n}$.  

A more natural and non-perturbative approach to estimating the field shift involves calculating the difference in level energies derived from two distinct calculations, employing separate sets of parametrizations to describe the nuclear charge distribution.
%Considering that the field shift effect exists for both the electronic and the nuclear states, 
In our case, the energy difference between clock transitions in $^{229}\text{Th}^{3+}$ and $^{229}\text{Th}^{4+}$ can be determined by performing four independent calculations on the atomic energy of the electronic ground state ($E_\text{g}$) in $^{229}\text{Th}^{3+}$,$^{229m}\text{Th}^{3+}$,$^{229}\text{Th}^{4+}$ and $^{229m}\text{Th}^{4+}$. 
Consequently, the energy difference of the nuclear transition is obtained as given by
\begin{equation}\label{eq_En}
\begin{aligned}
\Delta E_{N} = & (E_{\text{g},^{229m}\text{Th}^{3+}}-E_{\text{g},^{229}\text{Th}^{3+}}) \\
&-(E_{\text{g},^{229m}\text{Th}^{4+}} - E_{\text{g},^{229}\text{Th}^{4+}}).
\end{aligned}
\end{equation}
%Here, the first term represents the field shift in the ground state of Th VI induced by the nuclear ground and isomeric states, while the second term corresponds to the same effect in Th V. Consequently, their difference reflects the nuclear transition energy shift between Th IV and Th V.

However, as noted by Grant in~\cite{Grant1980}, \textit{ab initio} estimation of field shift by computing energy difference between levels from self-consistent field (SCF)  calculation is inherently flawed if the numerical calculation is not adequately converged. This approach demands extremely high-precision \textit{ab initio} atomic structure calculations to extract the subtle difference in $E_g$ under nearly identical nuclear potentials. Moreover, performing four independent calculations with equally high precision is significantly more time-consuming compared to the perturbation method.
Thanks to the development of our relativistic atomic structure package, GRASPG~\cite{si2024}—an optimized and enhanced version of the GRASP2018 package~\cite{Fischer.2019.V237.p184}—we can now perform significantly larger-scale multiconfiguration Dirac-Hartree-Fock (MCDHF) calculations with greater efficiency, leading to improved computational convergence.

%The calculation was carried out using the relativistic atomic structure package GRASPG based on the configuration state function generators (CSFGs)~\cite{si2024}, which is a further optimized and improved version of the GRASP2018 package~\citep{Fischer.2019.V237.p184} and relevant improvements can be found in~\cite{Li.2023.V11.p12,LI.2023.V283.p108562}
The electronic wavefunctions describing the electronic states of the atom, referred to
as atomic state functions (ASFs), are expressed as expansions over $\rm N_{CSF}$
configuration state functions (CSFs) within the framework of the multiconfiguration methods~\cite{Fischer.2016.V49.p182004}, characterized
by total angular momentum $J$, the total magnetic quantum number $M_J$ and parity $P$:
\begin{equation}\label{MCDF01}
\centering
\Psi(\Gamma J M_J)= \sum_{j=1}^{N_{CSF}} c_{j}\Phi(\gamma_{j} J M_J),
\end{equation}
where $\gamma_{j}$ represents the configuration, coupling, and other quantum numbers
necessary to uniquely describe the CSFs.

The wavefunctions for both valence and core electrons of $^{229(m)}$Th$^{4+}$ and $^{229(m)}$Th$^{3+}$ were determined independently.
The radial parts of the Dirac orbitals, along with the mixing coefficients $c_j$, were obtained in the MCDHF calculations. 
In this procedure, the Hamiltonian is represented by the Dirac-Coulomb Hamiltonian:
\begin{equation}
H_{\rm DC}= \sum_{i=1}^N(c~\bm{\alpha_i}\cdot\bm{p_i}+(\beta_i-1)c^2+V_i)+\sum_{i<j}^N\frac{1}{r_{ij}},
\end{equation}
where $\bm{\alpha}$ and $\beta$ are the $4\times4$ Dirac matrices and $c$ is the speed of light in atomic units, 
$V_i$ is the potential from an extended Fermi nuclear charge distribution~\cite{PARPIA1996}, 
\begin{equation}
\rho^n(r)=\frac{\rho_0}{1+e^{(r-c)/a}}
\end{equation}
where $\rho^{n}(r)$ is the nuclear radial charge density, $\rho_0$ is a constant.
The parameter $a$ is related to the skin thickness as
$t=(4\text{ln}3)a$, where the default skin thickness $t=2.30$~fm~\cite{PARPIA1996} was applied for both $^{229}$Th and $^{229m}$Th.
The recommended rms radius of $^{229}$Th, 5.7557~fm~\cite{Angeli2013}, and the rms radius difference $\delta \langle r^2 \rangle ^{229\text{m}, 229} = 0.01085(130)$~$\rm{fm^2}$, which is derived by averaging $\delta \langle r^2 \rangle ^{229\text{m}, 229}=$0.0105(13)~$\rm{fm^2}$ and 0.0112(13)~$\rm{fm^2}$ reported in~\cite{Safronova2018} and in~\cite{PhysRevLett.131.263002}, were used to determine the other parameter $c$.
Due to the difference of the rms radii between $^{229}$Th and $^{229m}$Th, their nuclear charge distributions and nuclear potentials also vary, which causes a slight shift in the electronic energy levels.

As an initial step, Dirac-Hartree-Fock calculations were performed for the single reference configurations:  $6s^26p^6$ for $^{229(m)}$Th$^{4+}$ and $6s^26p^65f$ for $^{229(m)}$Th$^{3+}$, respectively. 
Subsequently, in the MCDHF procedure, the CSFs expansions for both $^{229(m)}$Th$^{4+}$ and $^{229(m)}$Th$^{3+}$ were generated by allowing single and double substitutions from all the subshells outside the [Kr]-core of the reference configurations, to an active set (AS).
The active set was incrementally expanded to $n\leq10,l\leq5$ (AS$_{10}$, labeled by its maximum principle quantum number) layer by layer to monitor the convergence.
During this process, orbitals from previous sets were kept frozen, and only the outermost orbital of each symmetry in the newly added orbital set was optimized.
The transverse photon (Breit) interaction and the leading quantum electrodynamic (QED) corrections (vacuum polarization and self-energy) can be accounted for in subsequent
relativistic configuration interaction (RCI) calculations.
In the RCI calculations, the Dirac
orbitals from the MCDHF calculations are fixed, and only the mixing coefficients of the CSFs are determined by diagonalizing the Hamiltonian matrix.

The maximum number of CSFs in RCI calculations for $^{229(m)}$Th$^{4+}$ and $^{229(m)}$Th$^{3+}$ are 58,779 and 2,066,564, respectively.
By taking the advantage that the spin-angular integration is independent of the principal quantum numbers of the orbitals, the introduction of CSFG makes it possible to infer the spin-angular coefficients for a group of interacting CSFs from a relatively small number of CSFGs~\cite{LI.2023.V283.p108562}.
For example, the 58,779 and 2,066,564 CSFs can be expanded by only 4,890 and 192,999 CSFGs, significantly reducing both execution time and memory requirement.

%To further reduce the computation load, we utilized a prior condensation method~\cite{LI.2023.V283.p108562}, based on the fact that the electron correlation effects can be addressed by only a fraction of the CSFs, since many of them with small expansion coefficients contribute negligibly to the total energy and the computed expectation values. This approach involves performing an initial RCI calculation based on the CSFGs of AS$_{7}$, retaining only the CSFGs whose squared weights cumulatively account for 99.99999\% of the total weight. This resulting condensed set of CSFGs was then used for subsequent RCI calculations involving larger orbital sets, from AS$_{8}$ to AS$_{10}$. With the aid of the prior condensation method, the number of CSFs/CSFGs can be reduced by a factor of 2, while the computed energy shift and field shit parameters deviated from those obtained using full-interaction calculations by only 1.5~ppm.

\setlength{\tabcolsep}{8pt} %  6pt
\begin{table}[ht!]
\caption{Calculated energy difference ($\Delta E_\text{N}$ in atomic unit) and frequency shift ($\Delta \nu_\text{N}$ in MHz) of nuclear transition frequency between $^{229}$Th$^{3+}$ and $^{229}$Th$^{4+}$.}
\begin{tabular}{lcccccccccccccccrrr}
\hline
%     &   & \multicolumn{3}{c}{Condesed}\\
%     \cline{3-5}
 AS$_{n}$ &  $\Delta E_\text{N}$ & $\Delta \nu_\text{N}$ \\
\hline
AS$_7$  &  $-$9.82E$-$08 & $-646$ \\
AS$_8$  &  $-$9.75E$-$08 & $-642$  \\
AS$_9$  &  $-$9.72E$-$08 & $-640$   \\
AS$_{10}$ &  $-$9.71E$-$08 & $-639$ \\

\hline
\end{tabular}\label{tab-nu}
\end{table}

\setlength{\tabcolsep}{8pt} %  6pt
\begin{table}[ht!]
\caption{The difference in the squared large component of the radial amplitudes for ${\rm ^{229}Th^{3+}}$ and ${\rm ^{229}Th^{4+}}$ at the first grid point of the radius away from zero.}
\begin{tabular}{lcccccccccccccccrrr}
\hline
n$s$  & $P(1)^2_{\rm ^{229}Th^{3+}}-P(1)^2_{\rm ^{229}Th^{4+}}$ \\
\hline
$1s $ & 4.01634E$-$17\\ 
$2s $ & 7.94241E$-$17\\ 
$3s $ & 6.85517E$-$18\\ 
$4s $ & $-$6.60226E$-$17\\ 
$5s $ & $-$2.06146E$-$16\\ 
$6s $ & $-$8.05703E$-$16\\ 
\hline
\end{tabular}\label{tab-Pr}
\end{table}

%We completed the calculations on the atomic energy of the electronic ground state for ${\rm ^{229}Th^{3+}}  $,  ${\rm ^{229m}Th^{3+}}  $,  ${\rm ^{229}Th^{4+}}$ and ${\rm ^{229m}Th^{4+}}$. 
Based on four calculations of the ground state electronic energy levels of  $^{229(m)}$Th$^{4+}$ and $^{229(m)}$Th$^{3+}$, the frequency difference of the nuclear transition between $^{229}$Th$^{4+}$ and $^{229}$Th$^{3+}$ can be calculated to be $-639(1)$~MHz, as listed in Table~\ref{tab-nu} (see the energy of individual states and the computing convergency in the \textit{appendix}). The uncertainty  arises from computational convergency, defined as the deviation to the result using the penultimate AS$_n$ . 
Since the electronic effects, including electron correlation, Breit, and QED contributions were treated in the same way for $^{229(m)}$Th$^{4+}$ and $^{229(m)}$Th$^{3+}$, their contributions largely canceled out, resulting in a highly precise determination of the energy difference between them.
%Note that the convergence uncertainty in calculating the field shift difference is significantly smaller than that for individual atomic energy levels. 
%(\textcolor{red}{it is better to give a reason why the uncertainty in new method is smaller, note that the uncertainty of energy difference is much smaller than the energy shift it for Th3+ or Th4+, we need to give a reason.}) The reason is that all the electronic effect such as correlation, Breit and leading QED are mostly canceled in the identical electronic structure.

We also employed the perturbation theory to obtain the frequency difference. Considering only the first term,
%using the calculated first-term of field shift constants for ${\rm ^{229}Th^{3+}}$ and ${\rm ^{229}Th^{4+}}$ 
%and $\delta \langle r^2 \rangle ^{229\text{m}, 229} = 0.01085(130)$~$\rm{fm^2}$, which is derived by averaging 0.0105(13)~$\rm{fm^2}$ and 0.0112(13)~$\rm{fm^2}$ reported in ~\cite{Safronova2018} and in ~\cite{PhysRevLett.131.263002}, respectively. 
combining the rms difference of $\delta \langle r^2 \rangle ^{229\text{m}, 229} = 0.01085(130)$~$\rm{fm^2}$ with our calculated $\Delta F_{0} = -57.1$~$\rm GHz/fm^2$ using the RIS4 program~\cite{EKMAN2019433}, leads to the frequency shift to be $-620$~MHz, which is 19 MHz smaller than the one we get from the energy differences.  
The deviation can be attributed to the omission of the other terms arising from the nuclear deformation in the calculations. 
Assuming this discrepancy mainly originates from the second term $\Delta F_{2} \delta \langle r^{4} \rangle ^{229,229m}$, and utilizing our calculated $\Delta F_{2} = 0.074$~$\rm GHz/fm^4$, we extract the $\delta \langle r^{4} \rangle ^{229,229m}$ to be 0.260~$\rm{fm^4}$.

Duba and Flambaum~\cite{PhysRevLett.131.263002} have also calculated electron-induced shift between the nuclear transition frequencies of  $^{229}$Th$^{3+}$ and $^{229}$Th$^{4+}$ using the SD+CI and RPA method, obtaining $\Delta F_0=-55.0~\rm GHz/fm^2$, where the same [Rn] electronic core was assumed for both ions. 
In contrast, in our approach, we generated the electronic wave functions for each ion independently, thus core polarization by the additional 5f electron is taken into account. 
As shown in table~\ref{tab-Pr}, the presence of the $5f$ electron causes a slight change in the large component $P(r)^2$ of the radial wave function in $^{229}$Th$^{3+}$ compared to $^{229}$Th$^{4+}$. 
The overall impact can be estimated by considering the contribution of the n$s$ electrons, which leads to a lower electron density at the nucleus. This results in a reduced $F$ value for $^{229}$Th$^{3+}$ compared to the case where it shares the same electronic core as $^{229}$Th$^{4+}$. 

In order to exam the effect of electronic core on the nuclear transition frequency, we performed an additional  calculation on $^{229}$Th$^{3+}$ with the identical [Rn] electronic core of $^{229}$Th$^{4+}$, resulting in the $\Delta F_0= -51.4 ~\rm GHz/fm^2$. Thus, the variation of electronic core effect is about 10\%, corresponding to 62 MHz in the absolute frequency shift. Although this effect is not significant compared to the total frequency shift, it exceeds the expected uncertainty of the nuclear clock transition by more than 11 orders of magnitude. The deviation between our result ($-51.4~\rm GHz/fm^2$) and the value ($-55.0~\rm GHz/fm^2$) reported in~\cite{PhysRevLett.131.263002} can be attributed to the uncertainty arising from different theoretical methods. 

To derive the nuclear clock transition based on single trapped $^{229}\text{Th}^{3+}$ from the latest precise measurement of $^{229}\text{Th}^{4+}$ in the crystal by Zhang et. al.~\cite{Nature2024}, our study solves the major effect induced by the adding electron. However, due to the electromagnetic multipole moments of the nucleus, the transition frequency can be affected by the hyperfine interaction~\cite{Zhang2024229ThF4TF, Rellergert2010}. For the Th-doped crystal system, the dominant effect arises from the electromagnetic environment in the crystal. Zhang et. al.~\cite{Nature2024} has successfully resolved the hyperfine splitting resulting from the coupling between the nuclear quadrupole moment and the electric field gradient inside the CaF$_2$ crystal. Additionally, the electric monopole term, resulting from the interaction between the nuclues and its surrounding electron cloud in the lattice, induces a frequency shift of less than 100~MHz and a broadening on the order of 10~kHz/K~\cite{Rellergert2010}. The coupling between the nuclear magnetic moment and the magnetic field generated by the other nuclei contributes less than 10~kHz~\cite{Rellergert2010} to line broadening. 

As a result, using the experimental transition frequency of $^{229}\text{Th}^{4+}$, 2,020,407,384,335(2)~kHz, reported in ~\cite{Nature2024}, we determine the nuclear transition frequency of isolated $^{229}\text{Th}^{3+}$ to be $2,020,406,745 (1)_\text{comp.}(77)_{\delta \langle r^2 \rangle} (100)_\text{ext.}$~MHz. The first uncertainty comes from our MCDHF computations. The second one is caused by the error of the rms radii difference from the literature. The last error arises from the influence of the crystal environment, as discussed above.

For an isolated $^{229}$Th$^{3+}$ ion, 
the hyperfine structure arises from the interaction between the 
the unpaired $5f$ valence electron and the nuclear electromagnetic multipole moments.
The nuclear $I$ and electronic $J$ angular momenta couple to a total momentum
${F} = {I} + {J}$.
To first-order, the hyperfine energies are expressed in terms of the hyperfine interaction constants $A$ and $B$ that are related to the nuclear magnetic dipole moment $\mu$, electric quadrupole moment $Q$ and nuclear spin $I$~\cite{PhysRev.97.380}.

The $A$ and $B$ for the nuclear ground state have been measured as 82.2(6) and 2269(6)~MHz~\cite{PhysRevLett.106.223001}, respectively. The ratios of the magnetic dipole and electric quadrupole moments between the ground and isomer nuclear states have been determined to be $\mu_{is}/\mu_{g}=-1.04(15)$ and $Q_{is}/Q_{g}=0.57003(1)$~\cite{Nature2018,Nature2024}. 
Therefore, the $A$ and $B$ for the nuclear isomer state can be calculated to be $-142(21)$ and 1293(7)~MHz, respectively.
Table~\ref{tab-HFS} lists the hyperfine energy shifts relative to their centroid nuclear transition frequencies. The primary source of uncertainty stems from the inaccuracies in the reference magnetic dipole moments.

%The hyperfine splitting is around 1 ppm of the nuclear transition frequency, which is $13-14$ orders of magnitude larger than the uncertainty of the nuclear clock.
%Although the first-order contribution dominates the hyperfine splitting, the second-order contribution can not be omitted compared to the high accuracy of the nuclear clock.
%By adding the hyperfine energy shift to the centroid of the nuclear transition frequency of $^{229}$Th$^{3+}$, we determine the 12 hyperfine frequencies of nuclear clock transition for \textbf{}Th$^{3+}$.

\begin{table}
\caption{Hyperfine energy shifts in nuclear transition frequency of $^{229}\text{Th}^{3+}$ (in MHz) relative to their centroid.}
\begin{tabular}{lcrcccccccccccccrrr} 
\hline
  Isomer & Ground & Energy shift \\
\hline
$F=1$ &	$F=0$	 & 784(117)  \\
$F=1$ &	$F=1$	 & 1247(115) \\
$F=1$ &	$F=2$	 & 1967(113) \\
$F=2$ &	$F=1$	 & $-73(71)$  \\
$F=2$ &	$F=2$	 & 648(68)  \\
$F=2$ &	$F=3$	 & 1218(69) \\
$F=3$ &	$F=2$	 & -362(9) \\
$F=3$ &	$F=3$	 & 209(10)  \\
$F=3$ &	$F=4$	 & 16(11)   \\
$F=4$ &	$F=3$	 & 673(80) \\
$F=4$ &	$F=4$	 & 480(81) \\
$F=4$ &	$F=5$	 & $-1292(80)$   \\
\hline
\end{tabular}\label{tab-HFS}
\end{table}

In conclusion, we calculated the nuclear clock transition frequency shift between $^{229}\text{Th}^{3+}$ and $^{229}\text{Th}^{4+}$ using the non-perturbative MCDHF theory. The effect from nuclear charge distributions of the nuclear ground and isomeric states are considered in our calculation. Compared with the leading order approximation in field shift calculations, we extract the $ \delta \langle r^{4} \rangle ^{229,229m}$. Moreover, we investigate the impact of variations in the electronic core wave function on the nuclear transition frequency shift induced by the adding 5f electron, which accounts for approximately 10\% of the total shift. These findings refine previous calculations, providing essential insights for future precision spectroscopy and nuclear clock development.

\section{Acknowledgment}
R.S., C.S. and C.C. acknowledge the support by National Key Research and Development Program of China under Grant No. 2022YFA1602303, 2022YFA1602500. B.T. acknowledges the support by National Key Research and Development Program of China under Grant No.2023YFA1606501, 2022YFA1602504 and the National Natural Science Foundation of China under contract No. 12147101 and 12204110. X.J.K. and N.X. acknowledge the support by the National Key Research and Development Program of China under Grant No. 2024YFA1610900 and the National Natural Science Foundation of China under Contract No. 12447106 and 12147101. Y.G.M. thanks the National Natural Science Foundation of China under Contract No. 12147101 and the Guangdong Major Project of Basic and Applied Basic Research No. 2020B0301030008.

\section{Appendix}

\iffalse
\setlength{\tabcolsep}{6pt} % 
\begin{table*}[ht!]
\caption{Calculated leading (in $\rm GHz/fm^2$) and second order (in $\rm GHz/fm^4$) field shift electronic factors for $F_{\rm ^{229}Th^{4+}}  $ and  $F_{\rm ^{229}Th^{3+}}  $ with precision up to 8 and 7 significant digits respectively, and their difference $\Delta F_0 = F_{0, \rm ^{229}Th^{3+}} - F_{0,\rm ^{229}Th^{4+}}$ and $\Delta F_2 = F_{2,\rm ^{229}Th^{3+}} - F_{2,\rm ^{229}Th^{4+}}$, as functions of active sets.}
\begin{tabular}{lcccccccccccccccrrr}
\hline
%     &   & \multicolumn{3}{c}{Condesed}\\
%     \cline{3-5}
 ASs & $F_{0,\rm ^{229}Th^{4+}}  $&  $F_{0, \rm ^{229}Th^{3+}}  $&  $\Delta F_0$ & $F_{2,\rm ^{229}Th^{4+}}$&  $F_{2,\rm ^{229}Th^{3+}}$ &  $\Delta F_2$   \\
\hline
AS7  & 2306748.8& 2306696.5 & -53.3 &-2973.179 & -2973.110 &  0.0688  \\
AS8  & 2306738.6& 2306678.3 & -60.3 &-2973.166 & -2973.088 &  0.0779   \\
AS9  & 2306724.3& 2306667.7 & -56.6 &-2973.147 & -2973.074 &  0.0731   \\
AS10 & 2306720.9& 2306663.8 & -57.1 &-2973.143 & -2973.069 &  0.0738  \\
\hline
\end{tabular}\label{tab-FS}
\end{table*}
\fi

\setlength{\tabcolsep}{6pt} % 
\begin{table*}[ht!]
\caption{Calculated energy of the atomic ground state for $\rm ^{229(m)}Th^{3+}$ and $\rm ^{229(m)}Th^{4+}$ 
%in their ground and isomeric states (in atomic unit) 
%with precision up to 15 significant digits, 
and the difference $\Delta E_{\rm g,\rm Th \, IV} = E_{\rm g, \rm ^{229m}Th^{3+}} - E_{\rm g,\rm ^{229}Th^{3+}}$ and $\Delta E_{\rm g, \rm Th \, V} = E_{\rm g,\rm ^{229m}Th^{4+}} - E_{\rm g,\rm ^{229}Th^{4+}}$ , as functions of active sets.}
\begin{tabular}{lcccccccccccccccrrr}
\hline
%     &   & \multicolumn{3}{c}{Condesed}\\
%     \cline{3-5}
 ASs & $E_{\rm g,\rm ^{229}Th^{3+}}  $&  $E_{\rm g, \rm ^{229m}Th^{3+}}  $&  $\Delta E_{\rm g, \rm Th \, VI}$ & $E_{\rm g,\rm ^{229}Th^{4+}}$&  $E_{\rm g,\rm ^{229m}Th^{4+}}$ &  $\Delta E_{\rm g,\rm Th \, V}$   \\
\hline
AS7  & -26452.1231423864 & -26452.1196339200 & 0.0035084664  &-26451.1352841710 & -26451.1317756064 &  0.0035085646   \\
AS8  & -26452.2466260137 & -26452.2431175449 & 0.0035084688  &-26451.2059508799 & -26451.2024423136 &  0.0035085663    \\
AS9  & -26452.2894648398 & -26452.2859563704 & 0.0035084694  &-26451.2365257811 & -26451.2330172145 &  0.0035085666    \\
AS10 & -26452.3073192168 & -26452.3038107472 & 0.0035084696  &-26451.2517689877 & -26451.2482604210 &  0.0035085667   \\
\hline
\end{tabular}\label{tab-FS}
\end{table*}

\bibliographystyle{aip}
\bibliography{reference}% 

\end{document}